\documentclass[11pt]{article}
\usepackage[english]{babel}
\begin{document}

\date{}
\title{\textbf{BRST structure of non-linear superalgebras}}
\author{\textsc{M.~Asorey}${}^{a}$\thanks{E-mail: asorey@saturno.unizar.es},
\textsc{P.M.~Lavrov}${}^{a,b}$\thanks{ E-mail: lavrov@tspu.edu.ru},
\textsc{O.V.~Radchenko}${}^{b}$\thanks{E-mail: radchenko@tspu.edu.ru} and
\textsc{A.~Sugamoto}${}^{c}$\thanks{E-mail: sugamoto@phys.ocha.ac.jp}\\
\\${}^{a}$\textit{Departamento de F\'{\i}sica Te\'{o}rica,}
\\\textit{Facultad de Ciencias Universidad de Zaragoza,}
\\\textit{50009 Zaragoza, Spain}\\\\
${}^{b}$\textit{Department of Mathematical Analysis,}
\\\textit{Tomsk State Pedagogical University,}
\\\textit{634041 Tomsk, Russia}\\
\\${}^{c}$\textit{Department of Physics, Ochanomizu University,}
\\\textit{Otsuka, Bunkyo-ku,}
\\\textit{Tokyo 112-8610, Japan}}
\maketitle

\begin{quotation}
In this paper we analyse the structure of the BRST charge of nonlinear superalgebras.
We consider quadratic non-linear superalgebras where a commutator (in terms of (super) Poisson brackets) of the generators is a quadratic polynomial of the
generators. We find  the explicit form of the BRST charge up to cubic order in Faddeev-Popov ghost fields for arbitrary quadratic nonlinear superalgebras.
We point out the existence of  constraints on structure constants of the superalgebra when the nilpotent BRST charge is  quadratic in Faddeev-Popov ghost fields. 
The general results are illustrated  by simple examples of superalgebras.
\end{quotation}

\section{Introduction}
The nilpotent BRST charge as the Noether charge of 
the global Becchi-Rouet-Stora-Tyutin supersymmetry \cite{brs,t} is a crucial element in both
Lagrangian \cite{BV} and Hamiltonian \cite{BFV} quantization methods of gauge theories (see also the reviews \cite{BF}). For general gauge theories, the existence theorem for the nilpotent BRST charge has been proven \cite{BF}. It proceeds by the construction of  the BRST charge by an infinite, in general, series expansion in the Faddeev-Popov ghost fields. Sometimes these series are truncated and reduce to  finite polynomials. The most remarkable examples are given by the Yang-Mills theories when the nilpotent BRST charge is a quadratic function of Faddeev-Popov ghost fields. Another interesting examples are given by some quadratic nonlinear Lie algebras \cite{SSvN,DH}. The interest on nonlinear algebras was initiated by discovery of conformal field theories
\cite{CFT} which led to a new class of gauge theories with the nonlinear gauge algebras, the so-called ${\cal W_N}$ algebras \cite{WN}. The BRST construction for such algebras was discussed in \cite{tm,SSvN}. This is  closely related to the problem of the BRST construction for quantum groups with quadratic nonlinear algebras \cite{QG}. Recently, it was shown that a special class of nonlinear gauge algebras arises in the Lagrangian BRST approach to higher spin theories on anti de Sitter
(AdS) space \cite{HSF}. Note that non-linear algebras of supersymmetry
arise for some quantum mechanical systems with periodic finite-gap potentials \cite{Plyushc}. 

The general analysis of the BRST structure given in \cite{SSvN,DH,BL} was restricted to quadratic nonlinear algebras. In this paper, we extend this analysis to the case of quadratic nonlinear superalgebras.

We make use of the standard definition of the Poisson superbracket in a 
phase space with coordinates $\Gamma=(Q^A,P_A),\;\epsilon(Q^A)=\epsilon(P_A)=
\epsilon_A$ ($\epsilon(X)$ denotes the Grassmann parity of a quantity $X$) for any two functions $F,G$
\begin{eqnarray}
\label{sPB} 
\{F,G\}=\frac{\partial F}{\partial Q^A}\frac{\partial G}{\partial P_A}-
\frac{\partial G}{\partial Q^A}\frac{\partial F}{\partial P_A}
(-1)^{\epsilon(F)\epsilon(G)}
\end{eqnarray}
where the derivatives with respect to momenta $P_A$ stand for left derivatives, and those with respect to corresponding coordinates $Q^A$ stand for  right derivatives. The Poisson superbracket (\ref{sPB}) obeys the following properties:

\noindent
(1) Generalized antisymmetry
\begin{eqnarray}
\label{antiPB} 
\{F,G\}=-(-1)^{\epsilon(F)\epsilon(G)}\{G,F\},
\end{eqnarray}
(2) Generalized Jacobi identity
\begin{eqnarray}
\label{JIPB} 
\{F,\{G,H\}\}(-1)^{\epsilon(F)\epsilon(H)}+cyclic\ perms.(F,G,H)\equiv 0,
\end{eqnarray}
(3) Grassmann parity
\begin{eqnarray}
\label{GrasPB} 
\epsilon(\{F,G\})=\epsilon(F)+\epsilon(G),
\end{eqnarray} 
(4) By-linearity
\begin{eqnarray}
\label{linPB} 
\{F+H,G\}=\{F,G\}+\{H,G\}, \quad (\epsilon(F)=\epsilon(H)),
\end{eqnarray}
(5) Leibniz rule
\begin{eqnarray}
\label{LeiPB} 
\{FH,G\}=F\{H,G\}+\{F,G\}H(-1)^{\epsilon(H)\epsilon(G)},\\
\nonumber
\{F,GH\}=\{F,G\}H+G\{F,H\}H(-1)^{\epsilon(F)\epsilon(G)}.
\end{eqnarray}

In the present paper, we study the nilpotent BRST charge for quadratic nonlinear superalgebras and find some special restrictions on structure constants when the nilpotent BRST charge is given in the simplest form. 

The paper is organized as follows. In Section 2  the Jacobi identities for quadratic nonlinear superalgebras are derived and some simple examples of such a kind of superalgebras are constructed. In Section 3 the classical nilpotent BRST charge for quadratic nonlinear superalgebras with  some special restrictions on structure constants is constructed. In Section 4 we consider some simple examples of superalgebras for which general approach can be applied. In Section 5 we present some concluding remarks.

\section{Nonlinear superalgebras}

Let us consider a phase space $M$ with local coordinates $\{(q^i,p_i), i=1,2,..,n; (\epsilon(q^i)=\epsilon(p_i)=\epsilon_i)\}$ and let 
$\{T_{\alpha}=T_{\alpha}(q,p),\;\epsilon(T_{\alpha})=\epsilon_{\alpha}\}$ be a set of independent functions on $M$.  We suppose that $T_{\alpha}$ satisfy 
the involution relations in terms of the Poisson superbracket
\begin{eqnarray}
\label{invT}
\{T_{\alpha},T_{\beta}\}=T_{\gamma}F^{\gamma}_{\alpha\beta}+
T_{\delta}T_{\gamma}V^{\gamma\delta}_{\alpha\beta},
\end{eqnarray}
where the Grassmann parities $\epsilon(F^{\gamma}_{\alpha\beta})=
\epsilon_\alpha+\epsilon_\beta+\epsilon_\gamma$,
$\epsilon(V^{\gamma\delta}_{\alpha\beta})=
\epsilon_\alpha+\epsilon_\beta+\epsilon_\gamma+\epsilon_\delta$
and structure constants $F^{\gamma}_{\alpha\beta}$ and
$V^{\gamma\delta}_{\alpha\beta}$ possess the symmetry properties
\begin{eqnarray}\label{q2}
F^{\gamma}_{\alpha\beta}=
-(-1)^{\epsilon_\alpha\epsilon_\beta}F^{\gamma}_{\beta\alpha},\quad
V^{\gamma\delta}_{\alpha\beta}=
-(-1)^{\epsilon_\alpha\epsilon_\beta}V^{\gamma\delta}_{\beta\alpha}=
(-1)^{\epsilon_\delta\epsilon_\gamma}V^{\delta\gamma}_{\alpha\beta}.
\end{eqnarray}
The Jacobi identities for (\ref{invT}) read
\begin{eqnarray}\label{q3}
&&F^{\mu}_{\alpha\sigma}F^{\sigma}_{\beta\gamma}
(-1)^{\epsilon_\alpha\epsilon_\gamma}+cyclic\ perms.(\alpha,\beta,\gamma)=0,\\
\label{q4}
&&\Big(V^{\mu\nu}_{\alpha\sigma}F^{\sigma}_{\beta\gamma}+
F^{\mu}_{\alpha\sigma}V^{\sigma\nu}_{\beta\gamma}
(-1)^{\epsilon_\alpha\epsilon_\nu}+
F^{\nu}_{\alpha\sigma}V^{\sigma\mu}_{\beta\gamma}
(-1)^{\epsilon_\mu(\epsilon_\alpha+\epsilon_\nu)}\Big)
(-1)^{\epsilon_\alpha\epsilon_\gamma}+cyclic\ perms.(\alpha,\beta,\gamma)=0,\\
\label{q5}
&&\Big(V^{\mu\nu}_{\alpha\sigma}V^{\sigma\lambda}_{\beta\gamma}
(-1)^{\epsilon_\lambda(\epsilon_\alpha+\epsilon_\mu)}
+cyclic\ perms.(\mu,\nu,\lambda)\Big)(-1)^{\epsilon_\alpha\epsilon_\gamma}
+cyclic\ perms.(\alpha,\beta,\gamma)=0.
\end{eqnarray}

The simplest case of superalgebras really involving fermionic functions is 
a superalgebra with three generators $T,G_1,G_2$ where $T$ is  bosonic ($\epsilon(T)=0$) and $G_1,G_2$ are fermionic ($\epsilon(G_1)=\epsilon(G_2)=1$). In particular, we have that $G_1^2=G_2^2=0$. The most general relations 
for the Poisson superbrackets of generators preserving the Grassmann parities 
have the form
\begin{eqnarray}
\label{exsPB}
\nonumber
&&\{T,G_1\}=a_1(T)G_1+a_2(T)G_2,\quad \{T,G_2\}=b_1(T)G_1+b_2(T)G_2,\\
\nonumber
&&\{G_1,G_1\}=\alpha_1(T)+\alpha_2(T)G_1G_2,\quad
\{G_2,G_2\}=\beta_1(T)+\beta_2(T)G_1G_2,\\
\nonumber
&&\{G_1,G_2\}=\gamma_1(T)+\gamma_2(T)G_1G_2.
\end{eqnarray} 
Here $a_i,b_i,\alpha_i,\beta_i,\gamma_i, i=1,2$ are polynomial functions of $T$. 
Jacobi identities for this algebra require the fulfilment of equations
\begin{eqnarray}
\label{eqsalg} 
\nonumber
&&\alpha^{'}_1a_1+\alpha_2\gamma_1=0, \quad \beta^{'}_1b_1+\beta_1\beta_2=0,\quad
\alpha^{'}_1a_2-\alpha_1\alpha_2=0, \quad \beta^{'}_1b_2+\beta_2\gamma_1=0,\\
\nonumber
&&a_1\gamma_1+a_2\beta_1+b_1\alpha_1+b_2\gamma_1=0,\quad
b_2^{'}a_1-a_1^{'}+a_2^{'}b_1+a_2b_1^{'}-b_1\alpha_2+a_2\beta_2=0,\\
\nonumber
&&2\gamma_1^{'}a_1+\alpha_1^{'}b_1+2\gamma_1\gamma_2+\alpha_2\beta_1=0,\quad
2\gamma_1^{'}a_2+\alpha_1^{'}b_2-2\gamma_2\alpha_1-\alpha_2\gamma_1=0,\\
\nonumber
&&\beta_1^{'}a_1+2\gamma_1^{'}b_1+\beta_2\gamma_1+2\gamma_2\beta_1=0,\quad
\beta_1^{'}a_2-2\gamma_1^{'}b_2-\alpha_1\beta_2-2\gamma_1\gamma_2=0,
\end{eqnarray}
where $f^{'}$ denotes the derivative of $f=f(T)$ with respect to $T$. We have  nine first order differential equations and one algebraic nonlinear equation with ten unknowns
$a_i,b_i,\alpha_i,\beta_i,\gamma_i, i=1,2$. We will not study the general solution
to this system and will just list below some special cases. We have the following 
examples:
\begin{eqnarray}
 \label{I}
&&1.\;\{T,G_1\}=0,\quad \{T,G_2\}=0,\quad \{G_1,G_1\}=\alpha(T),\\
\nonumber
&&\;\;\;\;\{G_2,G_2\}=\beta(T),\quad \{G_1,G_2\}=\gamma(T).\\
\nonumber
\\
 \label{II}
&&2.\;\{T,G_1\}=a(T)G_1,\quad \{T,G_2\}=a(T)G_2,\quad \{G_1,G_1\}=0,\\
\nonumber
&&\;\;\;\;\{G_2,G_2\}=\beta(T)G_1G_2,\quad \{G_1,G_2\}=\gamma(T)G_1G_2.\\
\nonumber
\\
\label{III}
&&3.\;\{T,G_1\}=a(T)G_2,\quad \{T,G_2\}=b(T)G_1,\quad \{G_1,G_1\}=\alpha(T)G_1G_2,\\
\nonumber
&&\;\;\;\;\{G_2,G_2\}=0,\quad \{G_1,G_2\}=\gamma(T)G_1G_2.\\
\nonumber
\\
 \label{IV}
&&4.\;\{T,G_1\}=a(T)G_1,\quad \{T,G_2\}=0,\quad\{G_1,G_1\}=\alpha(T)G_1G_2,\\
\nonumber
&&\;\;\;\;\{G_2,G_2\}=\beta(T)G_1G_2,\quad \{G_1,G_2\}=\gamma(T)G_1G_2.\\
\nonumber
\\
 \label{V}
&&5.\;\{T,G_1\}=0,\quad \{T,G_2\}=b_1(T)G_1+b_2(T)G_2,\quad \{G_1,G_1\}=0,
\\
\nonumber
 &&\;\;\;\;\{G_2,G_2\}=\beta(T)G_1G_2,\quad \{G_1,G_2\}=\gamma(T)G_1G_2.
\end{eqnarray}
If we restrict ourselves to the case of quadratic nonlinear superalgebras 
the examples (\ref{I})-(\ref{V}) of  superalgebras (\ref{invT}) reduce to
\begin{eqnarray}
 \label{Iq}
&&1.\;\alpha(T)=A_1T+A_2T^2,\quad \beta(T)=B_1T+B_2T^2,\quad 
\gamma(T)=D_1T+D_2T^2,\\
\nonumber
&&\;\;\; F^{1}_{22}=A_1,\quad F^{1}_{33}=B_1,\quad F^{1}_{23}=D_1,\quad 
V^{11}_{22}=A_2,\quad V^{11}_{33}=B_2,\quad V^{11}_{23}=\frac{1}{2}D_2.\\
\nonumber
\\
 \label{IIq}
&&2.\;a(T)=A_0+A_1T,\quad \beta(T)=B_0,\quad \gamma(T)=D_0,\\
\nonumber
&&\;\;\;F^2_{12}=A_0,\quad F^3_{13}=A_0,\quad V^{12}_{12}=\frac{1}{2}A_1,\quad
V^{13}_{13}=\frac{1}{2}A_1,\quad V^{23}_{33}=\frac{1}{2}B_0,\quad
V^{23}_{23}=\frac{1}{2}D_0 .\\
\nonumber
\\
\label{IIIq}
&&3.\;a(T)=A_0+A_1T,\quad b(T)=B_0+B_1T,\quad \alpha(T)=C_0,\quad \gamma(T)=D_0,\\
\nonumber
&&\;\;\;F^3_{12}=A_0,\quad F^2_{13}=B_0,\quad V^{13}_{12}=\frac{1}{2}A_1,\quad
V^{12}_{13}=\frac{1}{2}B_1,\quad V^{23}_{22}=\frac{1}{2}C_0,\quad
V^{23}_{23}=\frac{1}{2}D_0.\\
\nonumber
\\
 \label{IVq}
&&4.\;a(T)=A_0+A_1T,\quad \alpha(T)=C_0,\quad \beta(T)=B_0, \quad \gamma(T)=D_0,\\
\nonumber
&&\;\;\;F^2_{12}=A_0,\quad V^{12}_{12}=\frac{1}{2}A_1,\quad
V^{23}_{22}=\frac{1}{2}C_0,\quad V^{23}_{33}=\frac{1}{2}B_0,\quad
V^{23}_{23}=\frac{1}{2}D_0.\\
\nonumber
\\
 \label{Vq}
&&5.\;b_1(T)=B_0+B_1T,\quad b_2(T)=B_2+B_3T,\quad \beta(T)=B_4,\quad 
\gamma(T)=D_0,
\\
\nonumber
 &&\;\;\;F^2_{13}=B_0,\quad F^3_{13}=B_2,\quad V^{23}_{33}=\frac{1}{2}B_4,\quad
V^{12}_{13}=\frac{1}{2}B_1,\quad V^{13}_{13}=\frac{1}{2}B_3,\quad
V^{23}_{23}=\frac{1}{2}D_0.
\end{eqnarray} 
where we introduce the notation $T=T_1,G_1=T_2,G_2=T_3, \epsilon_1=0, \epsilon_2=\epsilon_3=1$.   
Note that in the example (\ref{I}) there are superalgebras which 
appear in quantum systems with periodic finite-gap potentials \cite{Plyushc}  
if we identify the Hamiltonian with $H=T$, the two supersymmetry generators 
${\cal Q}_1=G_1$, ${\cal Q}_2=G_2$ and $\gamma(T)=0, \alpha(T)=\beta(T)=
P_{2n+1}(H)$. The example (\ref{I}), (\ref{Iq}) contains the superalgebra 
for dynamical systems with Hamiltonian $H=T$ which is invariant under BRST $Q=G_1$ and anti-BRST ${\bar Q}=G_2$ symmetry (the canonical quantization method based on this supersymmetry was proposed in \cite{BLT}) if we identify $A_1=1, A_2=B_1=B_2=D_1=G_2=0$ :
$\{Q,Q\}=0, \{{\bar Q},{\bar Q}\}=0,\{H,Q\}=0, \{H,{\bar Q}\}=0, 
\{Q,{\bar Q}\}=H$. In the example (\ref{II}), (\ref{IIq})  there exists the so-called self-reproducing superalgebras (for self-reproducing algebras within BRST formalism see \cite{DH}). Indeed, in the example (\ref{II}) it is enough to choose $a(T)=A_0T, \beta(T)=0, \gamma(T)=D_0$
to get the self-reproducing superalgebra.

\section{BRST construction}

The main quantity in the generalized canonical formalism \cite{BV,BFV} for dynamical systems with the first class constraints 
$T_{\alpha}=T_{\alpha}(q,p), \epsilon(T_{\alpha})=\epsilon_{\alpha}$ 
fulfilling the property $\{T_{\alpha},T_{\beta}\}\approx 0$, where $\approx$ 
denotes equality on the surface $T_{\alpha}(q,p)=0$, is the BRST charge ${\cal Q}$.
Nonlinear superalgebras (\ref{invT}) belong to this class. 
The BRST charge require to introduce for each
constraint $T_{\alpha}$ an anti-commuting ghost $c^{\alpha}$ and an
anticommuting momenta ${\cal P}_{\alpha}$ having the following
 Grassmann parities
$\epsilon(c^{\alpha})=\epsilon({\cal P}_{\alpha})=\epsilon_\alpha+1$
and  ghost numbers $gh(c^{\alpha})=-gh({\cal P}_{\alpha})=1$. They  have to obey the relations
\begin{eqnarray}\label{q6}
 \{c^{\alpha},{\cal P}_{\beta}\}=\delta^{\alpha}_{\beta},\quad
 \{c^{\alpha},c^{\beta}\}=0,\quad
 \{{\cal P}_{\alpha},{\cal P}_{\beta}\}=0,\quad
 \{c^{\alpha},T_{\beta}\}=0,\quad \{{\cal P}_{\alpha},T_{\beta}\}=0.
\end{eqnarray}

The  BRST charge ${\cal Q}$ is defined as a solution to the
equation
\begin{eqnarray}\label{Q}
\{{\cal Q},{\cal Q}\}=0
\end{eqnarray}
which is an  odd function of the variables $(p,q,c,{\cal P})$, has ghost
number $gh({\cal Q})=1$ and satisfies the boundary condition
\begin{eqnarray}\label{q7}
\frac{\partial {\cal Q}}{\partial c^{\alpha}}\Big|_{c=0}=T_{\alpha}.
\end{eqnarray}
A solution to the problem can be obtained  in terms of power-series
expansions in the ghost variables
\begin{eqnarray}\label{q8}
{\cal Q}=T_{\alpha}c^{\alpha}+\sum_{k\geq 1}
{\cal P}_{\beta_k}\cdot\cdot\cdot{\cal P}_{\beta_2}{\cal
P}_{\beta_1}
U^{(k)\beta_1\beta_2..\beta_k}_{\;\;\;\alpha_1\alpha_2..\alpha_{k+1}}
c^{\alpha_{k+1}}\cdot\cdot\cdot c^{\alpha_2}c^{\alpha_1}
={\cal Q}_1+\sum_{k\geq 1}{\cal Q}_{k+1},
\end{eqnarray}
where the symmetry properties of $U^{(k)}$ in lower indices coincide with
the symmetries of monomials $c^{\alpha_{k+1}}c^{\alpha_k}
\cdot\cdot\cdot c^{\alpha_1}$ while in upper indices they are defined by 
the symmetries of ${\cal P}_{\beta_k}{\cal P}_{\beta_{k-1}}\cdot\cdot\cdot{\cal
P}_{\beta_1}$. In particular
\begin{eqnarray}
\nonumber 
U^{(k)\beta_1\beta_2..\beta_k}_{\;\;\;\alpha_1\alpha_2..\alpha_{k+1}}=
(-1)^{(\epsilon_{\alpha_1}+1)(\epsilon_{\alpha_2}+1)}
U^{(k)\beta_1\beta_2..\beta_k}_{\;\;\;\alpha_2\alpha_1..\alpha_{k+1}}=
(-1)^{(\epsilon_{\beta_1}+1)(\epsilon_{\beta_2}+1)}
U^{(k)\beta_2\beta_1..\beta_k}_{\;\;\;\alpha_1\alpha_2..\alpha_{k+1}}.
\end{eqnarray}

Let us now apply the BRST construction to nonlinear superalgebras (\ref{invT}).
In lower order, the nilpotency of  ${\cal Q}$  implies that 
\begin{eqnarray}\nonumber
{\cal P}_{\beta_1}\Big((-1)^{\epsilon_{\alpha_1}}[F^{\beta_1}_{\alpha_1\alpha_2}+
T_{\beta_2}V^{\beta_2\beta_1}_{\alpha_1\alpha_2}]-
2U^{(1)\beta_1}_{\alpha_1\alpha_2}\Big)c^{\alpha_2}c^{\alpha_1}=0.
\end{eqnarray}
Thus,  the structure function has to be of the form
$U^{(1)}$
\begin{eqnarray}\label{q10}
U^{(1)\gamma}_{\;\;\alpha\beta}=\frac{1}{2}\Big(F^{\gamma}_{\alpha\beta}+
T_{\delta}V^{\delta\gamma}_{\alpha\beta}\Big)(-1)^{\epsilon_\alpha},\quad
U^{(1)\gamma}_{\;\;\alpha\beta}=U^{(1)\gamma}_{\;\;\beta\alpha}
(-1)^{(\epsilon_{\alpha}+1)(\epsilon_{\beta}+1)}
\end{eqnarray}
and the contribution ${\cal Q}_2$ of  second order in ghosts $c^{\alpha}$ to
${\cal Q}$ is
\begin{eqnarray}\label{Q2}
{\cal
Q}_2=\frac{1}{2}{\cal P}_{\gamma}\Big(F^{\gamma}_{\alpha\beta}+
T_{\delta}V^{\delta\gamma}_{\alpha\beta}\Big)c^{\beta}c^{\alpha}
(-1)^{\epsilon_\alpha}.
\end{eqnarray}

Using Jacobi identities
(\ref{q3}), (\ref{q4}), (\ref{q5}), the 
condition of nilpotency for ${\cal Q}$ in the third order can be rewritten as
\begin{eqnarray}\label{q12}
(-1)^{\epsilon_{\beta_1}\epsilon_{\beta_2}}{\cal P}_{\beta_2}T_{\beta_1}
\Big(T_{\beta_3}V^{\beta_3\beta_2}_{\alpha_1\sigma}
V^{\sigma\beta_1}_{\alpha_2\alpha_3}
(-1)^{\epsilon_{\alpha_2}+
\epsilon_{\alpha_1}\epsilon_{\beta_1}}+
4U^{(2)\beta_2\beta_1}_{\;\;\alpha_1\alpha_2\alpha_3}
(-1)^{\epsilon_{\beta_2}}\Big)c^{\alpha_3}c^{\alpha_2}c^{\alpha_1}=0.
\end{eqnarray}
Let us introduce the following quantities
\begin{eqnarray}\label{X1}
&&X^{\beta_3\beta_2\beta_1}_{\alpha_1\alpha_2\alpha_3}=
V^{\beta_3\beta_2}_{\alpha_1\sigma}
V^{\sigma\beta_1}_{\alpha_2\alpha_3}
(-1)^{\epsilon_{\alpha_2}+
\epsilon_{\beta_1}\epsilon_{\alpha_1}},\\
\nonumber
&& X^{\beta_3\beta_2\beta_1}_{\alpha_1\alpha_2\alpha_3}=
X^{\beta_2\beta_3\beta_1}_{\alpha_1\alpha_2\alpha_3}
(-1)^{\epsilon_{\beta_2}\epsilon_{\beta_3}}=
X^{\beta_3\beta_2\beta_1}_{\alpha_1\alpha_3\alpha_2}
(-1)^{(\epsilon_{\alpha_2}+1)(\epsilon_{\alpha_3}+1)}
\end{eqnarray}
which define the nilpotency equation in the third order (\ref{q12}).
Symmetrization of this quantity with respect to lower indices can be done 
using the rule obtained in Appendix A (see (\ref{Xs3}))
\begin{eqnarray}
X^{\beta_3\beta_2\beta_1}_{[\alpha_1\alpha_2\alpha_3]}=
X^{\beta_3\beta_2\beta_1}_{\alpha_1\alpha_2\alpha_3}+
X^{\beta_3\beta_2\beta_1}_{\alpha_3\alpha_1\alpha_2}
(-1)^{(\epsilon_{\alpha_3}+1)(\epsilon_{\alpha_1}+\epsilon_{\alpha_2})}+
X^{\beta_3\beta_2\beta_1}_{\alpha_2\alpha_3\alpha_1}
(-1)^{(\epsilon_{\alpha_1}+1)(\epsilon_{\alpha_2}+\epsilon_{\alpha_3})}. 
\end{eqnarray}
Then the nilpotency condition (\ref{q12}) can be written  in the form
\begin{eqnarray}
(-1)^{\epsilon_{\beta_1}\epsilon_{\beta_2}}{\cal P}_{\beta_2}T_{\beta_1}
\Big(T_{\beta_3}X^{\beta_3\beta_2\beta_1}_{[\alpha_1\alpha_2\alpha_3]}+
12U^{(2)\beta_2\beta_1}_{\;\;\alpha_1\alpha_2\alpha_3}
(-1)^{\epsilon_{\beta_2}}\Big)c^{\alpha_3}c^{\alpha_2}c^{\alpha_1}=0. 
\end{eqnarray}
From the Jacobi identities (\ref{q5}) it follows that
\begin{eqnarray}
\label{JIX}
X^{\beta_3\beta_2\beta_1}_{[\alpha_1\alpha_2\alpha_3]}+
X^{\beta_1\beta_3\beta_2}_{[\alpha_1\alpha_2\alpha_3]}
(-1)^{\epsilon_{\beta_1}(\epsilon_{\beta_2}+\epsilon_{\beta_3})}+
X^{\beta_2\beta_1\beta_3}_{[\alpha_1\alpha_2\alpha_3]}
(-1)^{\epsilon_{\beta_3}(\epsilon_{\beta_1}+\epsilon_{\beta_2})}= 0.
\end{eqnarray}
Consider now the quantities $N^{\alpha}_{\alpha_1\alpha_2\alpha_3}$
\begin{eqnarray}\label{q13}
N^{\alpha}_{\alpha_1\alpha_2\alpha_3}=T_{\beta_1}T_{\beta_3}
X^{\beta_3\alpha\beta_1}_{[\alpha_1\alpha_2\alpha_3]}
(-1)^{\epsilon_{\alpha}\epsilon_{\beta_1}}.
\end{eqnarray}
Due to (\ref{JIX}) they satisfy the relations
\begin{eqnarray}\label{q14}
T_{\alpha}N^{\alpha}_{\alpha_1\alpha_2\alpha_3}=0.
\end{eqnarray}
Therefore,
$N^{\alpha}_{\alpha_1\alpha_2\alpha_3}$ 
can be rewritten in the form
\begin{eqnarray}\label{q15}
N^{\alpha}_{\alpha_1\alpha_2\alpha_3}=T_{\beta}
N^{\{\alpha\beta\}}_{\alpha_1\alpha_2\alpha_3},\quad N^{\{\alpha\beta\}}_{\alpha_1\alpha_2\alpha_3}=
-N^{\{\beta\alpha\}}_{\alpha_1\alpha_2\alpha_3}
(-1)^{\epsilon_\alpha\epsilon_\beta}.
\end{eqnarray}
Taking into account (\ref{q13}), (\ref{q15}) we can show that
$N^{\{\alpha\beta\}}_{\alpha_1\alpha_2\alpha_3}$ has a linear dependence on
$T_{\alpha}$ 
\begin{eqnarray}\label{q16}
N^{\{\alpha\beta\}}_{\alpha_1\alpha_2\alpha_3}=
T_{\sigma}N^{\{\alpha\beta\}\sigma}_{\alpha_1\alpha_2\alpha_3}.
\end{eqnarray}
In terms of these quantities the structure functions $U^{(2)}$ are given by
\begin{eqnarray}\label{3q2}
U^{(2)\beta_2\beta_1}_{\;\;\alpha_1\alpha_2\alpha_3}=-\frac{1}{12}
T_{\sigma} N^{\{\beta_2\beta_1\}\sigma}_{\alpha_1\alpha_2\alpha_3}
(-1)^{\epsilon_{\beta_2}+\epsilon_{\beta_1}\epsilon_{\beta_2}},\quad 
U^{(2)\beta_2\beta_1}_{\;\;\alpha_1\alpha_2\alpha_3}=
U^{(2)\beta_1\beta_2}_{\;\;\alpha_1\alpha_2\alpha_3}
(-1)^{(\epsilon_{\beta_1}+1)(\epsilon_{\beta_2}+1)}.
\end{eqnarray}
Using (\ref{q13}), (\ref{q15})
and the Jacobi identities (\ref{JIX}) we
obtain the following equations 
\begin{eqnarray}\label{w15}
N^{\{\beta_2\beta_1\}\beta_3}_{\alpha_1\alpha_2\alpha_3}+
N^{\{\beta_2\beta_3\}\beta_1}_{\alpha_1\alpha_2\alpha_3}
(-1)^{\varepsilon_{\beta_1}\varepsilon_{\beta_3}}=
X^{\beta_3\beta_2\beta_1}_{[\alpha_1\alpha_2\alpha_3]}+
X^{\beta_1\beta_2\beta_3}_{[\alpha_1\alpha_2\alpha_3]}
(-1)^{\epsilon_{\beta_1}\epsilon_{\beta_3}}
\end{eqnarray}
which  define an explicit form of
$N^{\{\alpha\beta\}\sigma}_{\alpha_1\alpha_2\alpha_3}$.
In particular the structure of (\ref{w15}) allows us to suggest the form of  
$N^{\{\beta_2\beta_1\}\beta_3}_{\alpha_1\alpha_2\alpha_3}$
\begin{eqnarray}\label{q18}
N^{\{\beta_2\beta_1\}\beta_3}_{\alpha_1\alpha_2\alpha_3}=
C^{\{\beta_2\beta_1\}\beta_3}_{\mu_1(\mu_2\mu_3)}
X^{\mu_3\mu_2\mu_1}_{[\alpha_1\alpha_2\alpha_3]}
\end{eqnarray}
where $C^{\{\beta_2\beta_1\}\beta_3}_{\mu_1(\mu_2\mu_3)}$ is a matrix
constructed from the unit matrices $\delta^{\alpha}_{\mu}$ obeying the following
symmetry properties 
\begin{eqnarray}
 \label{Csim}
C^{\{\beta_2\beta_1\}\beta_3}_{\mu_1(\mu_2\mu_3)}=
-C^{\{\beta_1\beta_2\}\beta_3}_{\mu_1(\mu_2\mu_3)}(-1)^{\beta_1\beta_2}=
C^{\{\beta_2\beta_1\}\beta_3}_{\mu_1(\mu_3\mu_2)}
(-1)^{\epsilon_{\mu_2}\epsilon_{\mu_3}}.
\end{eqnarray}
It is not difficult to find a general structure of
$C^{\{\beta_2\beta_1\}\beta_3}_{\mu_1(\mu_2\mu_3)}$ with the required
symmetry properties
\begin{eqnarray}\label{q19}
C^{[\beta_2\beta_1]\beta_3}_{\mu_1(\mu_2\mu_3)}&=&
C\Big(\delta^{\beta_2}_{\mu_1}\delta^{\beta_1}_{\mu_2}
\delta^{\beta_3}_{\mu_3}-
\delta^{\beta_2}_{\mu_1}\delta^{\beta_1}_{\mu_1}\delta^{\beta_3}_{\mu_3}
(-1)^{\epsilon_{\beta_1} \epsilon_{\beta_2}}+\\
\nonumber
&&+\delta^{\beta_2}_{\mu_1}\delta^{\beta_1}_{\mu_3}\delta^{\beta_3}_{\mu_2}
(-1)^{\epsilon_{\mu_2}\epsilon_{\mu_3}}-
\delta^{\beta_2}_{\mu_3}\delta^{\beta_1}_{\mu_1}\delta^{\beta_3}_{\mu_2}
(-1)^{\epsilon_{\beta_1}\epsilon_{\beta_2}+\epsilon_{\mu_2}\epsilon_{\mu_3}}\Big),
\end{eqnarray}
where $C$ is a constant. From (\ref{q18}) and (\ref{q19}) it follows that
\begin{eqnarray}
N^{\{\beta_2\beta_1\}\beta_3}_{\alpha_1\alpha_2\alpha_3}
=2C\Big(X^{\beta_3\beta_1\beta_2}_{[\alpha_1\alpha_2\alpha_3]}-
X^{\beta_3\beta_2\beta_1}_{[\alpha_1\alpha_2\alpha_3]}
(-1)^{\epsilon_{\beta_1}\epsilon_{\beta_2}} \Big).
\end{eqnarray}
Inserting  this result into (\ref{w15}) one obtains
\begin{eqnarray}\label{CX}
4CX^{\beta_3\beta_1\beta_2}_{[\alpha_1\alpha_2\alpha_3]}=
(2C+1)X^{\beta_3\beta_2\beta_1}_{[\alpha_1\alpha_2\alpha_3]}
(-1)^{\epsilon_{\beta_1}\epsilon_{\beta_2}}+
(2C+1)X^{\beta_1\beta_2\beta_3}_{[\alpha_1\alpha_2\alpha_3]}
(-1)^{\epsilon_{\beta_1}\epsilon_{\beta_3}+\epsilon_{\beta_2}\epsilon_{\beta_3}}.
\end{eqnarray}
Taking into account the relations (\ref{JIX}) and the symmetry of $X^{\beta_1\beta_2\beta_3}_{[\alpha_1\alpha_2\alpha_3]}$ we have
\begin{eqnarray}
 \nonumber
X^{\beta_1\beta_2\beta_3}_{[\alpha_1\alpha_2\alpha_3]}
(-1)^{\epsilon_{\beta_1}\epsilon_{\beta_3}+\epsilon_{\beta_2}\epsilon_{\beta_3}}=
-X^{\beta_3\beta_1\beta_2}_{[\alpha_1\alpha_2\alpha_3]}-
X^{\beta_3\beta_2\beta_1}_{[\alpha_1\alpha_2\alpha_3]}
(-1)^{\epsilon_{\beta_1}\epsilon_{\beta_2}}
\end{eqnarray}
and therefore one can rewrite (\ref{CX}) in the form
\begin{eqnarray}
\label{CXm} 
\Big(6C+1\Big)
X^{\beta_3\beta_1\beta_2}_{[\alpha_1\alpha_2\alpha_3]}=0
\end{eqnarray}
in full agreement with bosonic case \cite{BL}. 
Then we have two solutions to the nilpotency equation at cubic order in
ghost variables $c^{\alpha}$. 
In the first case there is no  restriction on the structure constants $V^{\alpha\beta}_{\gamma\delta}$ of a quadratic nonlinear algebra 
\begin{eqnarray}
\label{S1}  
C=-1/6.
\end{eqnarray}
It leads to
\begin{eqnarray}\label{SN}  
N^{\{\beta_2\beta_1\}\beta_3}_{\alpha_1\alpha_2\alpha_3}
=-\frac{1}{3}\Big(X^{\beta_3\beta_1\beta_2}_{[\alpha_1\alpha_2\alpha_3]}-
X^{\beta_3\beta_2\beta_1}_{[\alpha_1\alpha_2\alpha_3]}
(-1)^{\epsilon_{\beta_1}\epsilon_{\beta_2}} \Big).
\end{eqnarray}
Therefore
\begin{eqnarray}\label{3q2U}
U^{(2)\beta_2\beta_1}_{\;\;\alpha_1\alpha_2\alpha_3}=-\frac{1}{36}
T_{\beta_3} \Big(X^{\beta_3\beta_2\beta_1}_{[\alpha_1\alpha_2\alpha_3]}-
X^{\beta_3\beta_1\beta_2}_{[\alpha_1\alpha_2\alpha_3]}
(-1)^{\epsilon_{\beta_1}\epsilon_{\beta_2}} \Big)
(-1)^{\epsilon_{\beta_2}}
\end{eqnarray}
and
\begin{eqnarray}
\label{Q3} 
{\cal Q}_3=-\frac{1}{6}{\cal P}_1{\cal P}_2T_{\beta_3}
V^{\beta_3\beta_2}_{\alpha_1\sigma}
V^{\sigma\beta_1}_{\alpha_2\alpha_3}(-1)^{\epsilon_{\alpha_2}+\epsilon_{\beta_2}+
\epsilon_{\alpha_1}\epsilon_{\beta_1}}
c^{\alpha_3}c^{\alpha_2}c^{\alpha_1}.
\end{eqnarray}
The second possibility corresponds to restriction on structure constants of nonlinear
superalgebras 
\begin{eqnarray}\label{R1}
X^{\beta_3\beta_2\beta_1}_{[\alpha_1\alpha_2\alpha_3]}=0
\end{eqnarray}
or
\begin{eqnarray}
\label{ResV} 
V^{\beta_3\beta_2}_{\alpha_1\sigma}V^{\sigma\beta_1}_{\alpha_2\alpha_3}
(-1)^{\epsilon_{\alpha_1}(\epsilon_{\alpha_3}+\epsilon_{\beta_1})}
+cyclic\ perms.(\alpha_1,\alpha_2,\alpha_3)=0.
\end{eqnarray}
It means that $N^{\{\beta_2\beta_1\}\beta_3}_{\alpha_1\alpha_2\alpha_3}=0$
and  
\begin{eqnarray}\label{q23}
U^{(2)\beta_1\beta_2}_{\alpha_1\alpha_2\alpha_3}=0,\quad 
{\cal Q}_3=0.
\end{eqnarray}
In what follows we restrict ourselves to the case of superalgebras where the restrictions
(\ref{ResV}) are fulfilled. In that case the Jacobi identities (\ref{q5})
are satisfied. 

Now let us analyse  the constraints generated for the condition of nilpotency at  forth order
of ghost fields $c^{\alpha}$. It has the form
\begin{eqnarray}
\label{q160}\nonumber
&&(-1)^{\epsilon_{\beta_3}}{\cal P}_{\beta_3}{\cal P}_{\beta_2}T_{\beta_1}
\Big[\Big( F^{\beta_1}_{\gamma\sigma}+T_{\beta_4}
V^{\beta_4\beta_1}_{\gamma\sigma}\Big)
V^{\sigma\beta_2}_{\alpha_1\alpha_2}V^{\gamma\beta_3}_{\alpha_3\alpha_4}
(-1)^{\epsilon_{\alpha_1}+\epsilon_{\alpha_3}+\epsilon_{\beta_2}+
(\epsilon_{\alpha_1}+\epsilon_{\alpha_2})
\epsilon_{\beta_3}+\epsilon_{\gamma}(\epsilon_{\alpha_1}+\epsilon_{\alpha_2}+
\epsilon_{\beta_2})} \\
&&\;\;\;\;\;\;\;\;\;\;\;\;\;\;
+24U^{(3)\beta_1\beta_2\beta_3}_{\alpha_1\alpha_2\alpha_3\alpha_4}
\Big]c^{\alpha_4}c^{\alpha_3}c^{\alpha_2}c^{\alpha_1}=0.
\end{eqnarray}
Let us introduce the following quantities 
\begin{eqnarray}
\label{Y34} 
Y^{\beta_1\beta_2\beta_3}_{\alpha_1\alpha_2\alpha_3\alpha_4}&=&
F^{\beta_1}_{\gamma\sigma}V^{\sigma\beta_2}_{\alpha_1\alpha_2}
V^{\gamma\beta_3}_{\alpha_3\alpha_4}
(-1)^{\epsilon_{\alpha_1}+\epsilon_{\alpha_3}+\epsilon_{\beta_2}+
(\epsilon_{\alpha_1}+\epsilon_{\alpha_2})
\epsilon_{\beta_3}+\epsilon_{\gamma}(\epsilon_{\alpha_1}+\epsilon_{\alpha_2}+
\epsilon_{\beta_2})},\\
\label{X44} 
X^{\beta_4\beta_1\beta_2\beta_3}_{\alpha_1\alpha_2\alpha_3\alpha_4}&=&
V^{\beta_4\beta_1}_{\gamma\sigma}V^{\sigma\beta_2}_{\alpha_1\alpha_2}
V^{\gamma\beta_3}_{\alpha_3\alpha_4}
(-1)^{\epsilon_{\alpha_1}+\epsilon_{\alpha_3}+\epsilon_{\beta_2}+
(\epsilon_{\alpha_1}+\epsilon_{\alpha_2})
\epsilon_{\beta_3}+\epsilon_{\gamma}(\epsilon_{\alpha_1}+\epsilon_{\alpha_2}+
\epsilon_{\beta_2})}
\end{eqnarray}
which have the symmetry properties
\begin{eqnarray}
\label{sY34} 
&&Y^{\beta_1\beta_2\beta_3}_{\alpha_1\alpha_2\alpha_3\alpha_4}=
Y^{\beta_1\beta_2\beta_3}_{\alpha_2\alpha_1\alpha_3\alpha_4}
(-1)^{(\epsilon_{\alpha_1}+1)(\epsilon_{\alpha_2}+1)}=
Y^{\beta_1\beta_2\beta_3}_{\alpha_1\alpha_2\alpha_4\alpha_3}
(-1)^{(\epsilon_{\alpha_3}+1)(\epsilon_{\alpha_4}+1)}\\
\label{sX44} 
&&X^{\beta_4\beta_1\beta_2\beta_3}_{\alpha_1\alpha_2\alpha_3\alpha_4}=
X^{\beta_4\beta_1\beta_2\beta_3}_{\alpha_2\alpha_1\alpha_3\alpha_4}
(-1)^{(\epsilon_{\alpha_1}+1)(\epsilon_{\alpha_2}+1)}=
X^{\beta_4\beta_1\beta_2\beta_3}_{\alpha_1\alpha_2\alpha_4\alpha_3}
(-1)^{(\epsilon_{\alpha_3}+1)(\epsilon_{\alpha_4}+1)}.
\end{eqnarray}
In order to define the structure function $U^{(3)}$ correctly we have to
symmetrize quantities which appear in (\ref{Y34}) and (\ref{X44}) in the indices
$\alpha_1,\alpha_2,\alpha_3,\alpha_4$. Using the symmetrization (\ref{A1}), (\ref{A2}) (\ref{A3}) and of symmetry properties (\ref{sY34})  
and (\ref{sX44}) we get
\begin{eqnarray}
\label{NE4}
(-1)^{\epsilon_{\beta_3}}{\cal P}_{\beta_3}{\cal P}_{\beta_2}T_{\beta_1}
\Big(Y^{\beta_1\beta_2\beta_3}_{[\alpha_1\alpha_2\alpha_3\alpha_4]}+
T_{\beta_4}X^{\beta_4\beta_1\beta_2\beta_3}_{[\alpha_1\alpha_2\alpha_3\alpha_4]}
+144U^{(3)\beta_1\beta_2\beta_3}_{\alpha_1\alpha_2\alpha_3\alpha_4}
\Big)c^{\alpha_4}c^{\alpha_3}c^{\alpha_2}c^{\alpha_1}=0.
\end{eqnarray}
From (\ref{Y34}) and (\ref{X44}) and symmetry properties of structure constants
$F^{\alpha}_{\beta\gamma}, V^{\alpha\beta}_{\gamma\delta}$ it follows
\begin{eqnarray}
\label{sYX}
Y^{\beta_1\beta_2\beta_3}_{[\alpha_1\alpha_2\alpha_3\alpha_4]}=
Y^{\beta_1\beta_3\beta_2}_{[\alpha_1\alpha_2\alpha_3\alpha_4]}
(-1)^{(\epsilon_{\beta_2}+1)(\epsilon_{\beta_3}+1)},\quad
X^{\beta_4\beta_1\beta_2\beta_3}_{[\alpha_1\alpha_2\alpha_3\alpha_4]}=
X^{\beta_4\beta_1\beta_3\beta_2}_{[\alpha_1\alpha_2\alpha_3\alpha_4]}
(-1)^{(\epsilon_{\beta_2}+1)(\epsilon_{\beta_3}+1)} 
\end{eqnarray}
It is possible to show that
\begin{eqnarray}
\label{X44z} 
X^{\beta_4\beta_1\beta_2\beta_3}_{[\alpha_1\alpha_2\alpha_3\alpha_4]}=0
\end{eqnarray}
as consequence of restrictions (\ref{R1}) or (\ref{ResV}). Indeed, using definitions
(\ref{X44}) and restrictions (\ref{ResV}) we obtain the equations
\begin{eqnarray}
\nonumber
X^{\beta_4\beta_1\beta_2\beta_3}_{\alpha_1\alpha_2\alpha_3\alpha_4}&-&
V^{\beta_4\beta_1}_{\alpha_2\sigma}
V^{\sigma\beta_2}_{\alpha_1\gamma}
V^{\gamma\beta_3}_{\alpha_3\alpha_4}
(-1)^{\epsilon_{\alpha_1}+\epsilon_{\alpha_3}+\epsilon_{\beta_2}+
\epsilon_{\alpha_2}(\epsilon_{\alpha_1}+\epsilon_{\beta_2})+
\epsilon_{\beta_3}(\epsilon_{\alpha_1}+\epsilon_{\alpha_2})}
+\\
\label{EqX44} 
&+&V^{\beta_4\beta_1}_{\alpha_1\sigma}
V^{\sigma\beta_2}_{\alpha_2\gamma}
V^{\gamma\beta_3}_{\alpha_3\alpha_4}
(-1)^{\epsilon_{\alpha_1}+\epsilon_{\alpha_3}+\epsilon_{\beta_2}+
\epsilon_{\alpha_1}\epsilon_{\beta_2}+
\epsilon_{\beta_3}(\epsilon_{\alpha_1}+\epsilon_{\alpha_2})}=0.
\end{eqnarray}
Symmetrization of (\ref{EqX44}) in indices 
$\alpha_1, \alpha_2, \alpha_3, \alpha_4$ gives rise to
\begin{eqnarray}
\nonumber 
&&X^{\beta_4\beta_1\beta_2\beta_3}_{[\alpha_1\alpha_2\alpha_3\alpha_4]}-
V^{\beta_4\beta_1}_{\alpha_1\sigma}
X^{\sigma\beta_2\beta_3}_{[\alpha_2\alpha_3\alpha_4]}
(-1)^{(\epsilon_{\alpha_1}+1)(\epsilon_{\beta_2}+1)+
\epsilon_{\alpha_1}\epsilon_{\beta_3}}-\\
\label{SimX44} 
&&-V^{\beta_4\beta_1}_{\alpha_2\sigma}
V^{\sigma\beta_2}_{\alpha_1\gamma}
V^{\gamma\beta_3}_{\alpha_3\alpha_4}
(-1)^{\epsilon_{\alpha_1}+\epsilon_{\alpha_3}+\epsilon_{\beta_2}
\epsilon_{\alpha_2}(\epsilon_{\alpha_1}+\epsilon_{\beta_2})+
\epsilon_{\beta_3}(\epsilon_{\alpha_1}+\epsilon_{\alpha_2})}-
\\
\nonumber
&&-V^{\beta_4\beta_1}_{\alpha_2\sigma}
V^{\sigma\beta_2}_{\alpha_1\gamma}
V^{\gamma\beta_3}_{\alpha_3\alpha_4}
(-1)^{\epsilon_{\alpha_1}+\epsilon_{\alpha_2}+\epsilon_{\beta_2}
\epsilon_{\alpha_4}(\epsilon_{\alpha_1}+\epsilon_{\beta_2})+
\epsilon_{\beta_3}(\epsilon_{\alpha_1}+\epsilon_{\alpha_2})+
(\epsilon_{\alpha_4}+1)(\epsilon_{\alpha_2}+\epsilon_{\alpha_3})}-
\\
\nonumber
&&-V^{\beta_4\beta_1}_{\alpha_2\sigma}
V^{\sigma\beta_2}_{\alpha_1\gamma}
V^{\gamma\beta_3}_{\alpha_3\alpha_4}
(-1)^{\epsilon_{\alpha_1}+\epsilon_{\alpha_4}+\epsilon_{\beta_2}
\epsilon_{\alpha_3}(\epsilon_{\alpha_1}+\epsilon_{\beta_2})+
\epsilon_{\beta_3}(\epsilon_{\alpha_1}+\epsilon_{\alpha_3})+
(\epsilon_{\alpha_2}+1)(\epsilon_{\alpha_3}+\epsilon_{\alpha_4})}=0.
\end{eqnarray}
Multiplying these equations by $c^{\alpha_4}c^{\alpha_3}c^{\alpha_2}c^{\alpha_1}$
we have
\begin{eqnarray}
\label{XVc4} 
\Big(X^{\beta_4\beta_1\beta_2\beta_3}_{[\alpha_1\alpha_2\alpha_3\alpha_4]}-2
V^{\beta_4\beta_1}_{\alpha_1\sigma}
X^{\sigma\beta_2\beta_3}_{[\alpha_2\alpha_3\alpha_4]}
(-1)^{(\epsilon_{\alpha_1}+1)(\epsilon_{\beta_2}+1)+
\epsilon_{\alpha_1}\epsilon_{\beta_3}}\Big)
c^{\alpha_4}c^{\alpha_3}c^{\alpha_2}c^{\alpha_1}=0
\end{eqnarray}
or due to (\ref{R1})
\begin{eqnarray}
\label{Xc4} 
X^{\beta_4\beta_1\beta_2\beta_3}_{[\alpha_1\alpha_2\alpha_3\alpha_4]}
c^{\alpha_4}c^{\alpha_3}c^{\alpha_2}c^{\alpha_1}=0
\end{eqnarray}
which proves (\ref{X44z}).

Solutions of the nilpotency equation (\ref{NE4}) are given  by the quantities $Y^{\beta_1\beta_2\beta_3}_{[\alpha_1\alpha_2\alpha_3\alpha_4]}$ with symmetry properties (\ref{sYX}). We shall prove that these quantities satisfy the following symmetries 
\begin{eqnarray}
\label{sY12}  
Y^{\beta_1\beta_2\beta_3}_{[\alpha_1\alpha_2\alpha_3\alpha_4]}=
Y^{\beta_2\beta_1\beta_3}_{[\alpha_1\alpha_2\alpha_3\alpha_4]}
(-1)^{(\epsilon_{\beta_1}+1)(\epsilon_{\beta_2}+1)}.
\end{eqnarray}
In that case  we will have for $U^{(3)}$
\begin{eqnarray}
\label{U3}  
U^{(3)\beta_1\beta_2\beta_3}_{\alpha_1\alpha_2\alpha_3\alpha_4}=
-\frac{1}{144}Y^{\beta_1\beta_2\beta_3}_{[\alpha_1\alpha_2\alpha_3\alpha_4]}
\end{eqnarray}
and for contribution to BRST charge in the forth order
\begin{eqnarray}
\label{Q4}   
{\cal Q}_4=-\frac{1}{24}{\cal P}_{\beta_3}{\cal P}_{\beta_2}{\cal P}_{\beta_1}
F^{\beta_1}_{\gamma\sigma}V^{\sigma\beta_2}_{\alpha_1\alpha_2}
V^{\gamma\beta_3}_{\alpha_3\alpha_4}
(-1)^{\epsilon_{\alpha_1}+\epsilon_{\alpha_3}+\epsilon_{\beta_2}+
(\epsilon_{\alpha_1}+\epsilon_{\alpha_2})
\epsilon_{\beta_3}+\epsilon_{\gamma}(\epsilon_{\alpha_1}+\epsilon_{\alpha_2}+
\epsilon_{\beta_2})}
c^{\alpha_4}c^{\alpha_3}c^{\alpha_2}c^{\alpha_1}.
\end{eqnarray}
This result can be considered as a supersymmetric generalization of BRST charge in the
forth order for quadratic nonlinear Lie algebras \cite{SSvN}.

To prove (\ref{sY12}) we start with Jacobi identities (\ref{q4}) 
\begin{eqnarray}
\label{IJsc}  
\Big(V^{\beta_1\beta_2}_{\gamma\sigma}F^{\sigma}_{\alpha_1\alpha_2}
+F^{\beta_1}_{\gamma\sigma}V^{\sigma\beta_2}_{\alpha_1\alpha_2}
(-1)^{\epsilon_{\gamma}\epsilon_{\beta_2}}+
F^{\beta_2}_{\gamma\sigma}V^{\sigma\beta_1}_{\alpha_1\alpha_2}
(-1)^{\epsilon_{\beta_1}(\epsilon_{\gamma}+\epsilon_{\beta_2})}
\Big)(-1)^{\epsilon_{\gamma}\epsilon_{\alpha_2}}+cyclic\ perms.(\gamma,\alpha_1,\alpha_2)
=0.
\end{eqnarray}
Multiplying these equations from right by
\begin{eqnarray}
\nonumber 
V^{\gamma\beta_3}_{\alpha_3\alpha_4}(-1)^{\epsilon_{\alpha_1}+\epsilon_{\alpha_3}
+\epsilon_{\beta_2}+\epsilon_{\beta_3}(\epsilon_{\alpha_1+}\epsilon_{\alpha_2})+
\epsilon_{\alpha_1}\epsilon_{\gamma}},
\end{eqnarray}
and taking into account the definitions (\ref{X1}), (\ref{Y34}), we obtain
\begin{eqnarray}
\label{EqY}
\nonumber 
&&Y^{\beta_1\beta_2\beta_3}_{\alpha_1\alpha_2\alpha_3\alpha_4}-
Y^{\beta_2\beta_1\beta_3}_{\alpha_1\alpha_2\alpha_3\alpha_4}
(-1)^{(\epsilon_{\beta_1}+1)(\epsilon_{\beta_2}+1)}-\\
\nonumber
&&\;\;\;-F^{\beta_1}_{\alpha_2\sigma}
X^{\sigma\beta_2\beta_3}_{\alpha_1\alpha_3\alpha_4}
(-1)^{\epsilon_{\alpha_1}+\epsilon_{\beta_2}+\epsilon_{\alpha_2}
(\epsilon_{\beta_2}+\epsilon_{\beta_3})+\epsilon_{\alpha_1}
\epsilon_{\alpha_2}}-\\
\nonumber
&&\;\;\;-F^{\beta_2}_{\alpha_2\sigma}
X^{\sigma\beta_1\beta_3}_{\alpha_1\alpha_3\alpha_4}
(-1)^{\epsilon_{\alpha_1}+\epsilon_{\beta_2}+\epsilon_{\alpha_2}
(\epsilon_{\beta_1}+\epsilon_{\beta_3})+\epsilon_{\alpha_1}
\epsilon_{\alpha_2}+\epsilon_{\beta_1}\epsilon_{\beta_2}}+\\
&&\;\;\;+F^{\beta_1}_{\alpha_1\sigma}
X^{\sigma\beta_2\beta_3}_{\alpha_2\alpha_3\alpha_4}
(-1)^{\epsilon_{\alpha_1}+\epsilon_{\beta_2}+\epsilon_{\alpha_1}
(\epsilon_{\beta_2}+\epsilon_{\beta_3})}+\\
\nonumber
&&\;\;\;+F^{\beta_2}_{\alpha_1\sigma}
X^{\sigma\beta_1\beta_3}_{\alpha_2\alpha_3\alpha_4}
(-1)^{\epsilon_{\alpha_1}+\epsilon_{\beta_2}+\epsilon_{\alpha_1}
(\epsilon_{\beta_1}+\epsilon_{\beta_3})+\epsilon_{\beta_1}\epsilon_{\beta_2}}+\\
\nonumber
&&\;\;\;+V^{\beta_1\beta_2}_{\gamma\sigma}F^{\sigma}_{\alpha_1\alpha_2}
V^{\gamma\beta_3}_{\alpha_3\alpha_4}
(-1)^{\epsilon_{\alpha_1}+\epsilon_{\alpha_3}+\epsilon_{\beta_2}+
(\epsilon_{\beta_3}+\epsilon_{\gamma})(\epsilon_{\alpha_1}+\epsilon_{\alpha_2})}+\\
\nonumber
&&\;\;\;+V^{\beta_1\beta_2}_{\alpha_2\sigma}F^{\sigma}_{\gamma\alpha_1}
V^{\gamma\beta_3}_{\alpha_3\alpha_4}
(-1)^{\epsilon_{\alpha_1}+\epsilon_{\alpha_3}+\epsilon_{\beta_2}+
\epsilon_{\beta_3}(\epsilon_{\alpha_1}+\epsilon_{\alpha_2})+
\epsilon_{\alpha_1}(\epsilon_{\alpha_2}+\epsilon_{\gamma})}+\\
\nonumber
&&\;\;\;+V^{\beta_1\beta_2}_{\alpha_1\sigma}F^{\sigma}_{\alpha_2\gamma}
V^{\gamma\beta_3}_{\alpha_3\alpha_4}
(-1)^{\epsilon_{\alpha_1}+\epsilon_{\alpha_3}+\epsilon_{\beta_2}+
\epsilon_{\beta_3}(\epsilon_{\alpha_1}+\epsilon_{\alpha_2})}=0.
\end{eqnarray}
Multiplying these equation form the right by $c^{\alpha_4}c^{\alpha_3}c^{\alpha_2}c^{\alpha_1}$ and taking into account the symmetrization in indices $\alpha_1\alpha_2\alpha_3\alpha_4$, we get
\begin{eqnarray}
\label{EqYcs}
&&\Big[\frac{1}{6}
\Big(Y^{\beta_1\beta_2\beta_3}_{[\alpha_1\alpha_2\alpha_3\alpha_4]}-
Y^{\beta_2\beta_1\beta_3}_{[\alpha_1\alpha_2\alpha_3\alpha_4]}
(-1)^{(\epsilon_{\beta_1}+1)(\epsilon_{\beta_2}+1)}
\Big)+\\
\nonumber
&&\;\;\;+\frac{2}{3}\Big(F^{\beta_1}_{\alpha_1\sigma}
X^{\sigma\beta_2\beta_3}_{[\alpha_2\alpha_3\alpha_4]}
(-1)^{\epsilon_{\beta_1}+\epsilon_{\alpha_1}\epsilon_{\beta_2}}+
F^{\beta_2}_{\alpha_1\sigma}
X^{\sigma\beta_1\beta_3}_{[\alpha_2\alpha_3\alpha_4]}
(-1)^{\epsilon_{\beta_2}+\epsilon_{\alpha_1}\epsilon_{\beta_1}+
\epsilon_{\beta_1}\epsilon_{\beta_2}}
\Big)(-1)^{\epsilon_{\alpha_1}(1+\epsilon_{\beta_3})}+\\
\nonumber
&&\;\;\;+V^{\beta_1\beta_2}_{\gamma\sigma}F^{\sigma}_{\alpha_1\alpha_2}
V^{\gamma\beta_3}_{\alpha_3\alpha_4}
(-1)^{\epsilon_{\alpha_1}+\epsilon_{\alpha_3}+\epsilon_{\beta_2}+
(\epsilon_{\alpha_1}+\epsilon_{\alpha_2})(\epsilon_{\beta_3}+
\epsilon_{\gamma})}+\\
\nonumber
&&\;\;\;+2V^{\beta_1\beta_2}_{\alpha_1\sigma}F^{\sigma}_{\alpha_2\gamma}
V^{\gamma\beta_3}_{\alpha_3\alpha_4}
(-1)^{\epsilon_{\alpha_1}+\epsilon_{\alpha_3}+\epsilon_{\beta_2}+
(\epsilon_{\alpha_1}+\epsilon_{\alpha_2})\epsilon_{\beta_3}}
\Big]c^{\alpha_4}c^{\alpha_3}c^{\alpha_2}c^{\alpha_1}=0. 
\end{eqnarray}
or due to (\ref{R1})
\begin{eqnarray}
\label{EqYcsr}
&&\Big[\frac{1}{6}
\Big(Y^{\beta_1\beta_2\beta_3}_{[\alpha_1\alpha_2\alpha_3\alpha_4]}-
Y^{\beta_2\beta_1\beta_3}_{[\alpha_1\alpha_2\alpha_3\alpha_4]}
(-1)^{(\epsilon_{\beta_1}+1)(\epsilon_{\beta_2}+1)}
\Big)+\\
\nonumber
&&\;\;\;+V^{\beta_1\beta_2}_{\gamma\sigma}F^{\sigma}_{\alpha_1\alpha_2}
V^{\gamma\beta_3}_{\alpha_3\alpha_4}
(-1)^{\epsilon_{\alpha_1}+\epsilon_{\alpha_3}+\epsilon_{\beta_2}+
(\epsilon_{\alpha_1}+\epsilon_{\alpha_2})(\epsilon_{\beta_3}+
\epsilon_{\gamma})}+\\
\nonumber
&&\;\;\;+2V^{\beta_1\beta_2}_{\alpha_1\sigma}F^{\sigma}_{\alpha_2\gamma}
V^{\gamma\beta_3}_{\alpha_3\alpha_4}
(-1)^{\epsilon_{\alpha_1}+\epsilon_{\alpha_3}+\epsilon_{\beta_2}+
(\epsilon_{\alpha_1}+\epsilon_{\alpha_2})\epsilon_{\beta_3}}
\Big]c^{\alpha_4}c^{\alpha_3}c^{\alpha_2}c^{\alpha_1}=0. 
\end{eqnarray}
Consider now the Jacobi identities (\ref{q4}) 
\begin{eqnarray}
\nonumber
\Big(V^{\sigma\beta_3}_{\alpha_2\gamma}F^{\gamma}_{\alpha_3\alpha_4}
+F^{\sigma}_{\alpha_2\gamma}V^{\gamma\beta_3}_{\alpha_3\alpha_4}
(-1)^{\epsilon_{\alpha_2}\epsilon_{\beta_3}}+
F^{\beta_3}_{\alpha_2\gamma}V^{\gamma\sigma}_{\alpha_3\alpha_4}
(-1)^{\epsilon_{\sigma}(\epsilon_{\alpha_2}+\epsilon_{\beta_3})}
\Big)(-1)^{\epsilon_{\alpha_2}\epsilon_{\alpha_4}}+cyclic\ perms.(\alpha_2,\alpha_3,\alpha_4)
=0.
\end{eqnarray}
Multiplying these equation form the right by $c^{\alpha_4}c^{\alpha_3}c^{\alpha_2}c^{\alpha_1}$ and from left by
\begin{eqnarray}
\nonumber 
V^{\beta_1\beta_2}_{\alpha_1\sigma}(-1)^{\epsilon_{\alpha_1}+\epsilon_{\alpha_3}+
\epsilon_{\beta_2}+\epsilon_{\beta_3}\epsilon_{\alpha_1}+\epsilon_{\alpha_2}
\epsilon_{\alpha_4}},
\end{eqnarray}
we obtain
\begin{eqnarray}
\nonumber
&&\Big(V^{\beta_1\beta_2}_{\alpha_1\sigma}V^{\sigma\beta_3}_{\alpha_2\gamma}
F^{\gamma}_{\alpha_3\alpha_4}(-1)^{\epsilon_{\alpha_1}+\epsilon_{\alpha_3}+
\epsilon_{\beta_2}+\epsilon_{\beta_3}\epsilon_{\alpha_1}}+\\
\label{ConIJ1} 
&&+V^{\beta_1\beta_2}_{\alpha_1\sigma}F^{\sigma}_{\alpha_2\gamma}
V^{\gamma\beta_3}_{\alpha_3\alpha_4}
(-1)^{\epsilon_{\alpha_1}+\epsilon_{\alpha_3}+
\epsilon_{\beta_2}+\epsilon_{\beta_3}(\epsilon_{\alpha_1}+\epsilon_{\alpha_2})}
+\\
\nonumber
&&+V^{\beta_1\beta_2}_{\alpha_1\sigma}F^{\beta_3}_{\alpha_2\gamma}
V^{\gamma\sigma}_{\alpha_3\alpha_4}
(-1)^{\epsilon_{\alpha_1}+\epsilon_{\alpha_3}+
\epsilon_{\beta_2}+\epsilon_{\beta_3}\epsilon_{\alpha_1}+
\epsilon_{\sigma}(\epsilon_{\alpha_2}+
\epsilon_{\beta_3})}
\Big)c^{\alpha_4}c^{\alpha_3}c^{\alpha_2}c^{\alpha_1}
=0.
\end{eqnarray}
Notice that
\begin{eqnarray}
\label{ConAdd1} 
&&V^{\beta_1\beta_2}_{\alpha_1\sigma}F^{\beta_3}_{\alpha_2\gamma}
V^{\gamma\sigma}_{\alpha_3\alpha_4}
(-1)^{\epsilon_{\alpha_1}+\epsilon_{\alpha_3}+
\epsilon_{\beta_2}+\epsilon_{\beta_3}\epsilon_{\alpha_1}+
\epsilon_{\sigma}(\epsilon_{\alpha_2}+
\epsilon_{\beta_3})}
c^{\alpha_4}c^{\alpha_3}c^{\alpha_2}c^{\alpha_1}=\\
\nonumber
&&=-\frac{1}{3}X^{\beta_1\beta_2\gamma}_{[\alpha_1\alpha_3\alpha_4]}
F^{\beta_3}_{\gamma\alpha_2}
(-1)^{(\epsilon_{\gamma}+1)(\epsilon_{\alpha_1}+\epsilon_{\alpha_3}+
\epsilon_{\alpha_4})+\epsilon_{\beta_2}+\epsilon_{\alpha_1}\epsilon_{\beta_3}}
c^{\alpha_4}c^{\alpha_3}c^{\alpha_2}c^{\alpha_1}=0
\end{eqnarray}
due to (\ref{R1}). Then from (\ref{ConIJ1}) it follows
\begin{eqnarray}
\nonumber
&&\Big(V^{\beta_1\beta_2}_{\alpha_1\sigma}V^{\sigma\beta_3}_{\alpha_2\gamma}
F^{\gamma}_{\alpha_3\alpha_4}(-1)^{\epsilon_{\alpha_1}+\epsilon_{\alpha_3}+
\epsilon_{\beta_2}+\epsilon_{\beta_3}\epsilon_{\alpha_1}}+\\
\label{ConIJ2} 
&&+V^{\beta_1\beta_2}_{\alpha_1\sigma}F^{\sigma}_{\alpha_2\gamma}
V^{\gamma\beta_3}_{\alpha_3\alpha_4}
(-1)^{\epsilon_{\alpha_1}+\epsilon_{\alpha_3}+
\epsilon_{\beta_2}+\epsilon_{\beta_3}(\epsilon_{\alpha_1}+\epsilon_{\alpha_2})}
\Big)c^{\alpha_4}c^{\alpha_3}c^{\alpha_2}c^{\alpha_1}
=0.
\end{eqnarray}
Let us now consider some additional relations which can be derived from (\ref{ResV})
\begin{eqnarray}
\nonumber
V^{\beta_1\beta_2}_{\alpha_1\sigma}V^{\sigma\beta_3}_{\alpha_2\gamma}
(-1)^{\epsilon_{\alpha_1}(\epsilon_{\gamma}+\epsilon_{\beta_3})}+
V^{\beta_1\beta_2}_{\gamma\sigma}V^{\sigma\beta_3}_{\alpha_1\alpha_2}
(-1)^{\epsilon_{\gamma}(\epsilon_{\alpha_2}+\epsilon_{\beta_3})}+
V^{\beta_1\beta_2}_{\alpha_2\sigma}V^{\sigma\beta_3}_{\gamma\alpha_1}
(-1)^{\epsilon_{\alpha_2}(\epsilon_{\alpha_1}+\epsilon_{\beta_3})}=0 
\end{eqnarray}
Multiplying these equation form the right by
\begin{eqnarray}
\nonumber
F^{\gamma}_{\alpha_3\alpha_4}(-1)^{\epsilon_{\alpha_1}+\epsilon_{\alpha_3}+
\epsilon_{\beta_2}+\epsilon_{\gamma}\epsilon_{\alpha_1}}
c^{\alpha_4}c^{\alpha_3}c^{\alpha_2}c^{\alpha_1}
\end{eqnarray}
we obtain
\begin{eqnarray}
\label{ConAdd2}
&&\Big(2V^{\beta_1\beta_2}_{\alpha_1\sigma}V^{\sigma\beta_3}_{\alpha_2\gamma}
F^{\gamma}_{\alpha_3\alpha_4}
(-1)^{\epsilon_{\alpha_1}+\epsilon_{\alpha_3}+
\epsilon_{\beta_2}+\epsilon_{\beta_3}\epsilon_{\alpha_1}}+\\
\nonumber
&&+V^{\beta_1\beta_2}_{\gamma\sigma}V^{\sigma\beta_3}_{\alpha_1\alpha_2}
F^{\gamma}_{\alpha_3\alpha_4}
(-1)^{\epsilon_{\alpha_1}+\epsilon_{\alpha_3}+
\epsilon_{\beta_2}+\epsilon_{\gamma}(\epsilon_{\alpha_1}+
\epsilon_{\alpha_2}+\epsilon_{\beta_3})}
\Big)c^{\alpha_4}c^{\alpha_3}c^{\alpha_2}c^{\alpha_1}=0.
\end{eqnarray}
Now, we can take  into account the following relations
\begin{eqnarray}
\label{Rel1}
&&V^{\beta_1\beta_2}_{\gamma\sigma}V^{\sigma\beta_3}_{\alpha_1\alpha_2}
F^{\gamma}_{\alpha_3\alpha_4}
(-1)^{\epsilon_{\alpha_1}+\epsilon_{\alpha_3}+
\epsilon_{\beta_2}+\epsilon_{\gamma}(\epsilon_{\alpha_1}+
\epsilon_{\alpha_2}+\epsilon_{\beta_3})}
c^{\alpha_4}c^{\alpha_3}c^{\alpha_2}c^{\alpha_1}=\\
\nonumber
&&=-V^{\beta_1\beta_2}_{\sigma\gamma}F^{\gamma}_{\alpha_1\alpha_2}
V^{\sigma\beta_3}_{\alpha_3\alpha_4}
(-1)^{\epsilon_{\alpha_1}+\epsilon_{\alpha_3}+
\epsilon_{\beta_2}+(\epsilon_{\sigma}+\epsilon_{\beta_3})(\epsilon_{\alpha_1}+
\epsilon_{\alpha_2})}c^{\alpha_4}c^{\alpha_3}c^{\alpha_2}c^{\alpha_1}
\end{eqnarray}
and
\begin{eqnarray}
\label{Rel2}
&&V^{\beta_1\beta_2}_{\alpha_1\sigma}V^{\sigma\beta_3}_{\alpha_2\gamma}
F^{\gamma}_{\alpha_3\alpha_4}
(-1)^{\epsilon_{\alpha_1}+\epsilon_{\alpha_3}+
\epsilon_{\beta_2}+\epsilon_{\beta_3}\epsilon_{\alpha_1}}
c^{\alpha_4}c^{\alpha_3}c^{\alpha_2}c^{\alpha_1}=\\
\nonumber
&&=-V^{\beta_1\beta_2}_{\alpha_1\sigma}F^{\sigma}_{\alpha_2\gamma}
V^{\gamma\beta_3}_{\alpha_3\alpha_4}
(-1)^{\epsilon_{\alpha_1}+\epsilon_{\alpha_3}+
\epsilon_{\beta_2}+\epsilon_{\beta_3}(\epsilon_{\alpha_1}+\epsilon_{\alpha_2})}
c^{\alpha_4}c^{\alpha_3}c^{\alpha_2}c^{\alpha_1}.
\end{eqnarray}
which  can be  derived with the help of (\ref{ConIJ2}). From (\ref{ConAdd2}),
(\ref{Rel1}) and (\ref{Rel2}) it follows that
\begin{eqnarray}
\label{MaRel} 
&&\Big(V^{\beta_1\beta_2}_{\gamma\sigma}F^{\sigma}_{\alpha_1\alpha_2}
V^{\gamma\beta_3}_{\alpha_3\alpha_4}
(-1)^{\epsilon_{\alpha_1}+\epsilon_{\alpha_3}+\epsilon_{\beta_2}+
(\epsilon_{\alpha_1}+\epsilon_{\alpha_2})(\epsilon_{\beta_3}+
\epsilon_{\gamma})}+\\
\nonumber
&&\;\;\;+2V^{\beta_1\beta_2}_{\alpha_1\sigma}F^{\sigma}_{\alpha_2\gamma}
V^{\gamma\beta_3}_{\alpha_3\alpha_4}
(-1)^{\epsilon_{\alpha_1}+\epsilon_{\alpha_3}+\epsilon_{\beta_2}+
(\epsilon_{\alpha_1}+\epsilon_{\alpha_2})\epsilon_{\beta_3}}
\Big)c^{\alpha_4}c^{\alpha_3}c^{\alpha_2}c^{\alpha_1}=0.
\end{eqnarray}
Therefore we derive from (\ref{EqYcsr})
\begin{eqnarray}
\label{SymY4}  
\Big(Y^{\beta_1\beta_2\beta_3}_{[\alpha_1\alpha_2\alpha_3\alpha_4]}-
Y^{\beta_2\beta_1\beta_3}_{[\alpha_1\alpha_2\alpha_3\alpha_4]}
(-1)^{(\epsilon_{\beta_1}+1)(\epsilon_{\beta_2}+1)}\Big)
c^{\alpha_4}c^{\alpha_3}c^{\alpha_2}c^{\alpha_1}=0,
\end{eqnarray}
that proves the symmetry properties of $Y^{\beta_1\beta_2\beta_3}_{[\alpha_1\alpha_2\alpha_3\alpha_4]}$ (\ref{sY12}).\\

If we additionally require the fulfilment of restrictions on structure constants of superalgebra (\ref{invT})
\begin{eqnarray}
\label{AdRes}  
Y^{\beta_1\beta_2\beta_3}_{[\alpha_1\alpha_2\alpha_3\alpha_4]}=0
\end{eqnarray}
or
\begin{eqnarray}
\nonumber
\label{AdResc}  
&&F^{\beta_1}_{\gamma\sigma}V^{\sigma\beta_2}_{\alpha_1\alpha_2}
V^{\gamma\beta_3}_{\alpha_3\alpha_4}
(-1)^{\epsilon_{\alpha_1}+\epsilon_{\alpha_3}+\epsilon_{\beta_2}+
(\epsilon_{\alpha_1}+\epsilon_{\alpha_2})\epsilon_{\beta_3}+
\epsilon_{\gamma}(\epsilon_{\alpha_1}+\epsilon_{\alpha_2}+\epsilon_{\beta_2})}+\\
\nonumber
&&+F^{\beta_1}_{\gamma\sigma}V^{\sigma\beta_2}_{\alpha_2\alpha_3}
V^{\gamma\beta_3}_{\alpha_1\alpha_4}
(-1)^{\epsilon_{\alpha_1}+\epsilon_{\alpha_2}+\epsilon_{\beta_2}+
(\epsilon_{\alpha_2}+\epsilon_{\alpha_3})\epsilon_{\beta_3}+
(\epsilon_{\alpha_1}+1)(\epsilon_{\alpha_2}+\epsilon_{\alpha_3})+
\epsilon_{\gamma}(\epsilon_{\alpha_2}+\epsilon_{\alpha_3}+\epsilon_{\beta_2})}+\\
&&+F^{\beta_1}_{\gamma\sigma}V^{\sigma\beta_2}_{\alpha_3\alpha_1}
V^{\gamma\beta_3}_{\alpha_2\alpha_4}
(-1)^{\epsilon_{\alpha_2}+\epsilon_{\alpha_3}+\epsilon_{\beta_2}+
(\epsilon_{\alpha_1}+\epsilon_{\alpha_3})\epsilon_{\beta_3}+
(\epsilon_{\alpha_3}+1)(\epsilon_{\alpha_1}+\epsilon_{\alpha_2})+
\epsilon_{\gamma}(\epsilon_{\alpha_1}+\epsilon_{\alpha_3}+\epsilon_{\beta_2})}+\\
\nonumber
&&+F^{\beta_1}_{\gamma\sigma}V^{\sigma\beta_2}_{\alpha_2\alpha_4}
V^{\gamma\beta_3}_{\alpha_3\alpha_1}
(-1)^{\epsilon_{\alpha_2}+\epsilon_{\alpha_3}+\epsilon_{\beta_2}+
(\epsilon_{\alpha_4}+\epsilon_{\alpha_2})\epsilon_{\beta_3}+
(\epsilon_{\alpha_1}+1)(\epsilon_{\alpha_2}+\epsilon_{\alpha_3})
(\epsilon_{\alpha_4}+1)(\epsilon_{\alpha_1}+\epsilon_{\alpha_3})+
\epsilon_{\gamma}(\epsilon_{\alpha_2}+\epsilon_{\alpha_4}+\epsilon_{\beta_2})}+\\
\nonumber
&&+F^{\beta_1}_{\gamma\sigma}V^{\sigma\beta_2}_{\alpha_3\alpha_4}
V^{\gamma\beta_3}_{\alpha_1\alpha_2}
(-1)^{\epsilon_{\alpha_1}+\epsilon_{\alpha_3}+\epsilon_{\beta_2}+
(\epsilon_{\alpha_3}+\epsilon_{\alpha_4})\epsilon_{\beta_3}+
(\epsilon_{\alpha_1}+\epsilon_{\alpha_2})
(\epsilon_{\alpha_3}+\epsilon_{\alpha_4})+
\epsilon_{\gamma}(\epsilon_{\alpha_3}+\epsilon_{\alpha_4}+\epsilon_{\beta_2})}+\\
\nonumber
&&+F^{\beta_1}_{\gamma\sigma}V^{\sigma\beta_2}_{\alpha_1\alpha_4}
V^{\gamma\beta_3}_{\alpha_2\alpha_3}
(-1)^{\epsilon_{\alpha_1}+\epsilon_{\alpha_2}+\epsilon_{\beta_2}+
(\epsilon_{\alpha_1}+\epsilon_{\alpha_2})\epsilon_{\beta_3}+
\epsilon_{\alpha_4}+1)(\epsilon_{\alpha_2}+\epsilon_{\alpha_3})+
\epsilon_{\gamma}(\epsilon_{\alpha_1}+\epsilon_{\alpha_4}+\epsilon_{\beta_2})}=0,
\end{eqnarray}
then we can state that there exists  a nilpotent BRST charge in a canonical quadratic form
\begin{eqnarray}
\label{Qf} 
{\cal
Q}=T_{\alpha}c^{\alpha}+\frac{1}{2}{\cal P}_{\gamma}
\Big(F^{\gamma}_{\alpha\beta}+
T_{\delta}V^{\delta\gamma}_{\alpha\beta}\Big)c^{\beta}c^{\alpha}
(-1)^{\epsilon_\alpha}.
\end{eqnarray}
for any superalgebras (\ref{invT}) satisfying the additional restrictions 
(\ref{ResV}), (\ref{AdResc}) on its structure constants 
$F^{\alpha}_{\beta\gamma}$ and $V^{\alpha\beta}_{\gamma\delta}$.\\

\section{Simple examples}
In this Section we approach the construction of the nilpotent BRST charge of the form 
(\ref{Qf}) for the simple examples listed in Section 2. In what follows we will use the following
notation for  the ghost variables  $(c^1,c^2,c^3)=(c,\eta_1,\eta_2), ({\cal P}_1,{\cal P}_2,{\cal P}_3)=({\cal P}, P_1,P_2)$.\\

1. The explicit form of structure constants (\ref{Iq}) implies that the indices $\beta_1,\beta_2,\beta_3,\sigma$ of the non-trivial relations in restrictions (\ref{ResV}) must have the following values 
$\beta_1=\beta_2=\beta_3=\sigma=1$. Thus, the only relation that has to be verified is the vanishing of
\begin{eqnarray}
V^{11}_{\alpha_11}V^{11}_{\alpha_2\alpha_3}
(-1)^{\epsilon_{\alpha_1}\epsilon_{\alpha_3}}+ cyclic\ perms.(\alpha_1,\alpha_2,\alpha_3)
=0 
\end{eqnarray}
and this relation is satisfied because  $V^{11}_{\alpha 1}=0$. In order  to verify  (\ref{AdResc})  non-trivial relations must be satisfied when $\gamma=\sigma=1$ and all terms in these relations contain a factor $F^{\beta}_{11}=0$ which has to 
vanish. Therefore the nilpotent BRST charge for this example has to be of the form 
\begin{eqnarray}
\label{ExQ1} 
{\cal Q}&=&Tc+G_1\eta_1+G_2\eta_2+\frac{1}{2}A_1{\cal P}\eta_1^2+
\frac{1}{2}B_1{\cal P}\eta_2^2+
D_1{\cal P}\eta_1\eta_2+\\
\nonumber
&&+\frac{1}{2}A_2{\cal P}T\eta_1^2+
\frac{1}{2}B_2{\cal P}T\eta_2^2+\frac{1}{2}D_2{\cal P}T\eta_1\eta_2. 
\end{eqnarray}
In this example there are no restrictions on the parameters 
$(A_1,B_1,D_1,A_2,B_2,D_2 )$ which ultimately define superalgebras (\ref{Iq}).\\

2. The analysis of the relations (\ref{ResV}) require the following restrictions 
on structure constants of superalgebras (\ref{IIq})
\begin{eqnarray}
\label{AdResII} 
V^{12}_{12}=V^{13}_{13}=V^{23}_{23}=0, \quad (A_1=D_0=0).
\end{eqnarray}
Due to the vanishing of $V^{1\alpha}_{\beta\gamma}=0$ and $F^{1}_{\alpha\beta}=0$ for all values of $\alpha,\beta,\gamma$, the relations (\ref{AdResc}) are satisfied.  
The nilpotent BRST charge can in this case be written as
\begin{eqnarray}
\label{ExQ2} 
{\cal Q}=Tc+G_1\eta_1+G_2\eta_2+A_0(P_1\eta_1+P_2\eta_2)c+
\frac{1}{2}B_0(P_2G_1-P_1G_2)\eta_2^2.  
\end{eqnarray}
\\

3.  Analyzing the relations (\ref{ResV}) we obtain the following restrictions for the superalgebra (\ref{IIIq})
\begin{eqnarray}
\label{AdResIII} 
V^{12}_{13}=V^{13}_{12}=V^{23}_{22}=V^{23}_{23}=0, \quad (A_1=B_1=C_0=D_0=0),
\end{eqnarray} 
which reduces to a linear superalgebra
with the usual nilpotent BRST charge for linear superalgebras
\begin{eqnarray}
\label{ExQ3} 
{\cal Q}=Tc+G_1\eta_1+G_2\eta_2+A_0 P_2\eta_1 c+B_0P_1\eta_2 c.
\end{eqnarray}
\\

4. As  in the previous case, the  analysis of the relations (\ref{ResV}) imposes severe restrictions on the superalgebra (\ref{IVq})
\begin{eqnarray}
\label{AdResIV} 
V^{12}_{12}=V^{23}_{22}=V^{23}_{33}=V^{23}_{23}=0, \quad (A_1=B_0=C_0=D_0=0)
\end{eqnarray} 
which also reduces to a linear superalgebra.
The nilpotent BRST charge has the form 
\begin{eqnarray}
\label{ExQ4} 
{\cal Q}=Tc+G_1\eta_1+G_2\eta_2+A_0 P_1\eta_1 c.
\end{eqnarray}
\\

5. The analysis of the relations (\ref{ResV}) imposes  the following restrictions 
on structure constants of the superalgebras (\ref{Vq})
\begin{eqnarray}
\label{AdResV} 
V^{13}_{13}=V^{23}_{23}=V^{23}_{33}=0, \quad (B_3=D_0=B_4=0).
\end{eqnarray}
Imposing the vanishing of $V^{3\alpha}_{\beta\gamma}=0$ for all values of $\alpha,\beta,\gamma$ and 
$F^{2}_{13}\neq 0, F^{3}_{13}\neq 0$, the relations (\ref{AdResc}) are satisfied.  
The nilpotent BRST charge can then be written in the form
\begin{eqnarray}
\label{ExQ5} 
{\cal Q}=Tc+G_1\eta_1+G_2\eta_2+\Big(B_0 P_1 + B_2P_2+
\frac{1}{2}B_1({\cal P}G_1+P_1T)\Big)\eta_2c. 
\end{eqnarray}

\section{Discussion}
In this paper we have investigated the BRST structure of quadratic nonlinear
superalgebras of form (\ref{invT}) which are characterized by the structure constant
$F^{\alpha}_{\beta\gamma}$ and $V^{\alpha\beta}_{\gamma\delta}$. The explicit form of the BRST charge both in the second and third orders was found without any additional restrictions on the structure constants. In the case when the structure constants verify  the constraints (\ref{ResV}), the construction  of the BRST charge can be achieved up to the fourth
order in the ghost fields $c^{\alpha}$.  
We have found additional restrictions (see (\ref{AdResc})) on structure constants of any non-linear quadratic superalgebras when nilpotent BRST charge can be written in a canonical form (\ref{Qf}) which is quadratic in ghost fields $c^{\alpha}$. We have constructed simple quadratic nonlinear superalgebras with one bosonic and two fermionic generators and have verified all the constraints of the structure constants in order to explicitly construct  the BRST charge in the canonical form.

\section*{Acknowledgements}
The work of M.A. is partially supported by CICYT (grant FPA2006-2315)
and DGIID-DGA (grant2007-E24/2). P.M.L. acknowledges the MEC for the 
grant (SAB2006-0153). O.V.R. thanks Ochanomizu University for the financial support where the part of this work was done. The work of P.M.L. and O.V.R. was supported by
the grant for LRSS, project No.\ 2553.2008.2. The work of P.M.L.
was also supported  by the  RFBR-Ukraine grant No.\ 08-02-90490.

\vspace{.5cm}
\noindent{\Large{\bf Appendix}}
\hspace*{\parindent}
\vspace{0.5cm}
\begin{appendix}

\section{Symmetrization}
\renewcommand{\theequation}{\thesection.\arabic{equation}}
\setcounter{equation}{0}
\hspace*{\parindent}
\vspace{0.5cm}
\label{APP_A}

Let us now consider the procedure of symmetrization used for the  correct definition of structure
functions $U^{(k)}, k=2,3$. Let $X_{\alpha_1\alpha_2\alpha_3}$ be some quantities
appearing in expression 
$X=X_{\alpha_1\alpha_2\alpha_3}c^{\alpha_3}c^{\alpha_2}c^{\alpha_1}$. Due to known
symmetry properties of monomials $c^{\alpha_3}c^{\alpha_2}c^{\alpha_1}$, $X$ can be expressed in terms of $X_{[\alpha_1\alpha_2\alpha_3]}$ having required symmetry.
We have
\begin{eqnarray}
\nonumber
X_{[\alpha_1\alpha_2\alpha_3]}&=&
\frac{\partial^3 X}{\partial c^{\alpha_1}\partial c^{\alpha_2}
\partial c^{\alpha_3}}=\\
\nonumber
&=&X_{\alpha_1\alpha_2\alpha_3}
+X_{\alpha_3\alpha_1\alpha_2}
(-1)^{(\epsilon_{\alpha_1}+\epsilon_{\alpha_2})(\epsilon_{\alpha_3}+1)}+
X_{\alpha_2\alpha_3\alpha_1}(-1)^{(\epsilon_{\alpha_1}+1)(\epsilon_{\alpha_3}+
\epsilon_{\alpha_2})}+\\
\nonumber
&&+X_{\alpha_1\alpha_3\alpha_2}(-1)^{(\epsilon_{\alpha_2}+1)(\epsilon_{\alpha_3}+1)}
+X_{\alpha_2\alpha_1\alpha_3}(-1)^{(\epsilon_{\alpha_1}+1)(\epsilon_{\alpha_2 }+1)}
+\\
\nonumber
&&+X_{\alpha_3\alpha_2\alpha_1}
(-1)^{(\epsilon_{\alpha_1}+\epsilon_{\alpha_2})(\epsilon_{\alpha_3}+1)+
(\epsilon_{\alpha_1}+1)(\epsilon_{\alpha_2}+1)}=\\
\nonumber 
&=&\Big(X_{\alpha_1\alpha_2\alpha_3}
(-1)^{\epsilon_{\alpha_2}+\epsilon_{\alpha_1}\epsilon_{\alpha_3}}+
cyclic\ perms.(\alpha_1,\alpha_2,\alpha_3)\Big)
(-1)^{\epsilon_{\alpha_2}+\epsilon_{\alpha_1}\epsilon_{\alpha_3}}-\\
\nonumber
&&-\Big(X_{\alpha_1\alpha_3\alpha_2}
(-1)^{\epsilon_{\alpha_3}+\epsilon_{\alpha_1}\epsilon_{\alpha_2}}+
cyclic\ perms.(\alpha_1,\alpha_2,\alpha_3)\Big)
(-1)^{\epsilon_{\alpha_2}+\epsilon_{\alpha_2}(\epsilon_{\alpha_1}+
\epsilon_{\alpha_3})}
\end{eqnarray}
and
\begin{eqnarray}
\label{SymX}
 X=\frac{1}{3!}X_{[\alpha_1\alpha_2\alpha_3]}c^{\alpha_3}c^{\alpha_2}c^{\alpha_1}.
\end{eqnarray}
If $X_{\alpha_1\alpha_2\alpha_3}$ has additional symmetry properties
\begin{eqnarray}
X_{\alpha_1\alpha_2\alpha_3}=X_{\alpha_1\alpha_3\alpha_2}
(-1)^{(\epsilon_{\alpha_2}+1)(\epsilon_{\alpha_3}+1)}
\end{eqnarray} 
then
\begin{eqnarray}
\nonumber
X_{[\alpha_1\alpha_2\alpha_3]}&=&
2\Big(X_{\alpha_1\alpha_2\alpha_3}
+X_{\alpha_3\alpha_1\alpha_2}
(-1)^{(\epsilon_{\alpha_1}+\epsilon_{\alpha_2})(\epsilon_{\alpha_3}+1)}
+X_{\alpha_2\alpha_3\alpha_1}(-1)^{(\epsilon_{\alpha_1}+1)(\epsilon_{\alpha_2}+
\epsilon_{\alpha_3})}\Big)\\
\label{Xs3}
&=&
2\Big(X_{\alpha_1\alpha_2\alpha_3}
(-1)^{\epsilon_{\alpha_2}+\epsilon_{\alpha_1}\epsilon_{\alpha_3}}+
cyclic\ perms.(\alpha_1,\alpha_2,\alpha_3)\Big)(-1)^{\epsilon_{\alpha_2}+\epsilon_{\alpha_1}
\epsilon_{\alpha_3}}.
\end{eqnarray}
Let us now consider  quartic quantities in the ghost fields,
$Y=Y_{\alpha_1\alpha_2\alpha_3\alpha_4}
c^{\alpha_4}c^{\alpha_3}c^{\alpha_2}c^{\alpha_1}$. One can introduce 
the symmetric structure $Y_{[\alpha_1\alpha_2\alpha_3\alpha_4]}$
\begin{eqnarray}
 \nonumber
Y_{[\alpha_1\alpha_2\alpha_3\alpha_4]}=
\frac{\partial^4 Y}{\partial c^{\alpha_1}\partial c^{\alpha_2}
\partial c^{\alpha_3}\partial c^{\alpha_4}}
\end{eqnarray}
which can be expressed in terms of three indices symmetric quantities
\begin{eqnarray}
\nonumber
Y_{[\alpha_1\alpha_2\alpha_3\alpha_4]}&=&
Y_{\alpha_1[\alpha_2\alpha_3\alpha_4]}+Y_{\alpha_4[\alpha_1\alpha_2\alpha_3]}
(-1)^{(\epsilon_{\alpha_4}+1)(\epsilon_{\alpha_1}+\epsilon_{\alpha_2}+
\epsilon_{\alpha_3}+1)}+\\
\label{A1}
&&+Y_{\alpha_3[\alpha_4\alpha_1\alpha_2]}
(-1)^{(\epsilon_{\alpha_1}+\epsilon_{\alpha_2})(\epsilon_{\alpha_3}+
\epsilon_{\alpha_4})}+
Y_{\alpha_2[\alpha_3\alpha_4\alpha_1]}
(-1)^{(\epsilon_{\alpha_1}+1)(\epsilon_{\alpha_2}+\epsilon_{\alpha_3}+
\epsilon_{\alpha_4}+1)}. 
\end{eqnarray}
Then we have
\begin{eqnarray}
 Y=\frac{1}{4!}Y_{[\alpha_1\alpha_2\alpha_3\alpha_4]}
c^{\alpha_4}c^{\alpha_3}c^{\alpha_2}c^{\alpha_1},
\end{eqnarray}
and  if $Y_{\alpha_1\alpha_2\alpha_3\alpha_4}$ has additional symmetry properties
\begin{eqnarray}
\label{A2}
Y_{\alpha_1\alpha_2\alpha_3\alpha_4}=
Y_{\alpha_2\alpha_1\alpha_3\alpha_4}
(-1)^{(\epsilon_{\alpha_1}+1)(\epsilon_{\alpha_2}+1)}=
Y_{\alpha_1\alpha_2\alpha_4\alpha_3}
(-1)^{(\epsilon_{\alpha_3}+1)(\epsilon_{\alpha_4}+1)}, 
\end{eqnarray}
 one can finally show that 
\begin{eqnarray}
\label{A3}
&&Y_{[\alpha_1\alpha_2\alpha_3\alpha_4]}=
4\Big(Y_{\alpha_1\alpha_2\alpha_3\alpha_4}+
Y_{\alpha_2\alpha_3\alpha_1\alpha_4}
(-1)^{(\epsilon_{\alpha_1}+1)(\epsilon_{\alpha_2}+\epsilon_{\alpha_3})}+\\
\nonumber
&&\;\;\;+Y_{\alpha_3\alpha_1\alpha_2\alpha_4}
(-1)^{(\epsilon_{\alpha_3}+1)(\epsilon_{\alpha_1}+\epsilon_{\alpha_2})}+
Y_{\alpha_2\alpha_4\alpha_3\alpha_1}
(-1)^{(\epsilon_{\alpha_1}+1)(\epsilon_{\alpha_2}+\epsilon_{\alpha_3})+
(\epsilon_{\alpha_4}+1)(\epsilon_{\alpha_1}+\epsilon_{\alpha_3})}+\\
\nonumber
&&\;\;\;+Y_{\alpha_3\alpha_4\alpha_1\alpha_2}
(-1)^{(\epsilon_{\alpha_1}+\epsilon_{\alpha_2})
(\epsilon_{\alpha_3}+\epsilon_{\alpha_4})}+Y_{\alpha_1\alpha_4\alpha_2\alpha_3}
(-1)^{(\epsilon_{\alpha_4}+1)(\epsilon_{\alpha_2}+\epsilon_{\alpha_3})}\Big). 
\end{eqnarray}

\end{appendix}

\end{document}